\documentclass[twoside]{article}

\usepackage{PRIMEarxiv}

\usepackage[utf8]{inputenc} 
\usepackage[T1]{fontenc}    
\usepackage{hyperref}       
\usepackage{url}            
\usepackage{booktabs}       
\usepackage{amsfonts}       
\usepackage{nicefrac}       
\usepackage{microtype}      
\usepackage{lipsum}
\usepackage{fancyhdr}       
\usepackage{graphicx}       
\graphicspath{{media/}}     

\title{Scalable random number generation \\ for truncated log-concave distributions
}

\author{
  Michele Lambardi di San Miniato \\
  Department of Economics and Statistics \\
  University of Udine \\
  Udine, Italy \\
  \texttt{michele.lambardi@uniud.it} \\
   \And
  Euloge Clovis Kenne Pagui \\
  Department of Statistical Sciences \\
  University of Padova \\
  Padova, Italy \\
  \texttt{kenne@stat.unipd.it} \\
}

\usepackage{graphicx}
\usepackage{amsmath}
\usepackage{amssymb}
\usepackage{algorithm2e}

\renewcommand{\Pr}{\mathbb{P}}
\newcommand{\interv}{\mathcal{I}}
\newcommand{\support}{\mathcal{X}}
\newcommand{\trunc}[1]{{#1}_{\interv}}

\begin{document}
\maketitle

\begin{abstract}
Inverse transform sampling is an exceptionally general method to generate non-uniform-distributed random numbers, but can be rather unstable when simulating extremely truncated distributions. Many famous probability models share a property called log-concavity, which is not affected by truncation, so they can all be simulated via rejection sampling using Devroye's approach. This sampler is based on rejection and thus more stable than inverse transform, and uses a very simple envelope whose acceptance rate is guaranteed to be at least 20\%. The aim of this paper is threefold: firstly, to warn against the risk of wrongly simulating from truncated distributions; secondly, to motivate a more extensive use of rejection sampling to mitigate the issues; lastly, to motivate Devroye's automatic method as a practical standard in the case of log-concave distributions. We illustrate the proposal by means of simulations based on some Tweedie distributions, for their relevance in regression analysis.
\end{abstract}


\section{Introduction}

We consider the problem of non-uniform random number generation, when the aim is to simulate a univariate, log-concave, and extremely truncated probability distribution. Formally, let $X$ be a random variable with support $\mathcal{X}\subseteq\mathbb{R}$ and probability mass or density function $f(x)$, and let $\interv \subset \support$ be such that $\Pr(X\in\interv)>0$. The distribution is said \textit{log-concave} if $\log f(x)$ is a concave function. Moreover, the distribution of $X$ \textit{truncated to $\interv$} can be defined as the distribution of $X$ conditional to $X \in \interv$ and denoted by $\trunc{f}(x)$; henceforth, $f(x)$ is referred to as the \textit{base} distribution of $X$. It is possible to define $f_\interv(x)$ as
$$\trunc{f}(x) = f(x)/P(\interv) \,,\quad x \in \interv \,,$$
where $P(\interv)=\Pr(X \in \interv)$. Samples of $\trunc{f}(x)$ can be drawn and re-drawn from $f(x)$ until they lie in $\interv$, as per hit-or-miss. We consider the case when $\interv=\support \,\cap\, ]a,+\infty[$ for some $a \in \mathbb{R}$, but the software we provide supports truncation over intervals in the form $\interv=\support \, \cap \, ]a,b]$, as typical in the other \texttt{R} software.

The condition of interest is called ``extreme truncation" in the following and occurs when $P(\interv) \approx 0$, thus as $a$ approaches $\sup\support$. Extreme truncation makes hit-or-miss inefficient at simulating $\trunc{f}(x)$, because only a fraction $P(\interv)$ of samples from $f(x)$ is accepted on average, so a workaround is in order. The inverse transform method is a popular sampling algorithm, but it can be unreliable under extreme truncation \cite{Laud_2010}. Rejection sampling is more trustworthy \cite{Botev_2017} but cannot be deployed as straightforwardly in general. Devroye proposed an automatic rejection sampling method whose efficiency is uniformly lower-bounded over all the eligible distributions, which are the log-concave ones \cite{Devroye_1986,Devroye_1987}.

An idea flashes through one's mind when noticing that many classical distributions, including Tweedie ones, are log-concave and therefore so are truncations of these distributions \cite{Evans_2000}. In this paper, we analyze some risks in simulating truncated distributions via inverse transform. We show via simulations that Devroye's method can mitigate the issues at least with the eligible distributions, as it breaks down only under far more extreme truncation than inverse transform does. As a remark, the problem is not overcome at all, but only mitigated. We restrict the analysis to Tweedie distributions, which are of immediate relevance for regression analysis tasks, but our conclusions may also apply to other log-concave distributions as well.

We provide an \texttt{R} package\nocite{R} that implements Devroye's sampler for truncated log-concave distributions, not just Tweedie ones; the package only requires to specify a base probability mass or density function $f(x)$, the base cumulative distribution function $F(x)=\Pr(X\leq x)$, and a base mode function that returns the mode of the base distribution $m =\arg\sup_{x\in\support} f(x)$. All such functions can be consistently parameterized, like for instance the Poisson distribution with mean $\lambda>0$. Moreover, both $f(x)$ and $F(x)$ must be available in logarithmic form for meaningful evaluation in the case of extreme truncation.

The article is organized as follows. In Section \ref{sec:backg}, we review concepts like truncated and log-concave distributions. In Section \ref{sec:sampling}, we recall some additional definitions involved in the inverse transform and Devroye sampling methods. In Section \ref{sec:simulations}, a simulation study is presented for each implemented distribution. In Section \ref{sec:diag} we validate the diagnostic tools provided by our package. Conclusions are drawn in Section \ref{sec:discuss}.

\section{Background}\label{sec:backg}

Truncation is a common stratagem in statistical modeling of tail events. For instance, in risk prevention, the behavior of a system is analyzed conditional to rare and disruptive events \cite{Russell_1971,ergashev}. Related to truncation, though distinct, are censoring \cite{klein}, binning and quantization \cite{Cadez2002}. Each probability distribution comes with a built-in tail behavior that implies an assumption about censoring \cite{Nixon2004}. Truncated distributions are involved in process start-up, inventory control \cite{smith} and masking microdata \cite{parsa}. Truncation allows to expand modeling possibilities, by manipulating known distributions \cite{Tokmachev2018}. For instance, stochastic frontier analysis \cite{Aigner_1977,Meeusen_1977} involves positive-valued random variables, defined by truncating either normal, Laplace or other log-concave distributions \cite{Pitt_1981,Wheat_2019}. The truncated normal is especially popular \cite{thompson,Thomas2007,williams,chen}, followed by truncated binomial and Poisson distributions for zero-inflated modeling \cite{wenger}. Truncated distributions are also needed to assess some robust estimators \cite{Kunsch_1989}.

Extremely truncated distributions are especially challenging to simulate. In normal applications, stepping into them is considered a mistake, though, which motivates an ostrich approach. For instance, iterative sampling of Bayesian posterior distributions can be troubled by model misspecification or poor initialization \cite{Botev_2017}, so one can just step into parameter configurations that imply extremely truncated distributions, but some trial and error will suffice to get the analysis back on track. Other errors are more substantial than merely computational. The 2008 financial crisis was many sigma away from normality according to some light tail models in use at the time \cite{Nelson_2014}. Nonetheless, there are occasions when simulating extremely truncated distributions is a necessity more than a mistake. The software should be able to accommodate for legitimate prompts and make reliable simulations accordingly.

In applications, interest lies often on multivariate distributions that are analytically intractable. Simulation and Monte Carlo methods are just one option to speed up the analysis and sometimes even the only way out of it \cite{Rubinstein_2016}. Often, in Bayesian inference, the posterior distribution can be summarized only based on samples \cite{Botev_2017}. In addition, bootstrap is frequently resorted to in approximating standard errors associated with frequentist estimates \cite{Laan2001,Hong2009,Monchuk2010}. It is thus crucial that random number generators provide samples that reflect the desired distributions, but this aspect should not be taken for granted. In \texttt{R}\nocite{R}, many simulation-oriented packages \cite{truncdist,taskview} rely on the inverse transform sampling method, which fails at simulating extreme tail events \cite{Laud_2010}. Breakdowns can be due to inaccuracies \cite{Rmpfr} and algorithmic limitations \cite{Altman2003} that affect the evaluation of probability-related functions. Our aim here is to make it possible to carry out reliable simulations within standard finite-precision arithmetic, where the choice of algorithms is still crucial \cite{Loader2002}.

Truncated models are higher-dimensional than their base counterparts, since ruled by the additional parameters defining $\interv$, which makes diagnostics and prevention even more challenging. This is also why we restrict our discussion to truncation in the specific form used throughout this article, which is still relevant in practice. More complicated truncations can be defined in terms of mixtures. The scope of our discussion is further restricted, as we address the case of log-concave distributions. All such distributions have necessarily light, sub-exponential, tails \cite{Saumard2014}. This feature makes tail probabilities especially small and hard to evaluate meaningfully. Tweedie distributions, involved in the generalized linear model, are all log-concave; only gamma with shape parameter less than unit is not log-concave \cite{Philippe1997}, but it can be simulated indirectly based on a log-concave exponential power distribution \cite{Devroye_1986}. Log-concavity implies a concentration inequality and an envelope (or ``hat") distribution to perform rejection sampling, with execution time bounded in probability \cite{Devroye_1986}. This sampler is very general, can be tuned automatically, has acceptance rate greater or equal than 20\%, with eligible distributions, and is also numerically stable since based on rejection sampling \cite{Botev_2017}. The envelope is a mixture involving uniform and exponential distributions in the continuous case, uniform and geometric distributions in the discrete one, thus it is easy to simulate.

\section{Sampling truncated distributions}\label{sec:sampling}

In non-uniform random number generation \cite{Devroye_1986}, each sampling method needs some descriptor of the target distribution, like $f(x)$ in rejection sampling. Some descriptors can be troubled at times. For instance, consider the Laplace transform of the distribution \cite{Ridout_2008}: it works as a distribution descriptor, since in one-to-one correspondence with $f(x)$, which is a descriptor itself, but is numerically troubled when the support $\support$ is bounded (this problem is known as Gibbs phenomenon). Sampling methods based on the Laplace transform are expected to be unreliable in such occasions. Other issues may affect other distribution descriptors as well, thus in random number generation it is crucial to know which descriptor fails and \textit{when}. Inaccuracies and algorithmic limitations \cite{Altman2003}, which we broadly refer to as numerical issues, can affect many statistical tasks apart from random number generation \cite{McCullough1998,Melard2014}, but we address that issue in later sections.

Firstly, we recall few basic distribution descriptors for sampling tasks. In inverse transform sampling, the fundamental distribution descriptor is the quantile function, denoted by $q(p)$ and defined as the pseudo inverse of $F(x)$, that is,
$$q(p) = \inf\left\{x \in \mathbb{R} : F(x) \geq p\right\} \,,\quad p \in [0,1] \,.$$
When $X$ is an absolutely continuous random variable, it holds $q(p) = F^{-1}(p)$. The quantile function $q_\interv(p)$ of $X$ truncated to $\interv$ is defined as
$$\trunc{q}(p) = q\{F(a) + p \, P(\interv)\} \,.$$

\subsection{Inverse transform sampling}

If $q(p)$ is tractable, simulating a distribution can be based on the inverse transform sampling method (ITS). This technique can be relatively safe in normal applications. When considering extremely truncated distributions, though, this method may fail severely, to an extent that depends also on the specific implementation of the relevant distribution descriptor. This technique is nonetheless a standard one in the absence of more targeted samplers, due to its unrivaled generality. If $q(p)$ can be evaluated efficiently, and if a uniform-distributed random variable $U$ can be simulated, then also $X$ can be simulated, via ITS. One has to exploit the fact that
$$X \sim q(U) \,,$$
where ``$\sim$" denotes equivalence in distribution. The ITS will first generate samples of $U$ and then transform them via $q(p)$, in order to simulate samples of $X$.

For ITS to sample a truncated distribution, little modifications are needed, when $F(x)$ can be evaluated efficiently too. Consider $X$ being truncated to $\interv$. Then, $U$ can be transformed via $\trunc{q}(p)$ instead of $q(p)$, that is,
$$X \sim \trunc{q}(U) \,|\,X\in\interv \,.$$
ITS scales well \textit{in theory}, but the actual scalability depends on software implementations of $q(p)$ and $F(x)$. In \texttt{R} software, as of version 4.1.1, the package \texttt{stats} cannot evaluate standard normal quantiles beyond nine sigma. More extreme quantiles will be automatically imputed as $+\infty$. The package \texttt{truncnorm} \cite{truncnorm} instead imputes $q_\interv(p)\approx a$, which is more meaningful after concentration inequalities for log-concave distributions, as discussed later. Truncating the standard normal to $]+9,+\infty[$ is thus not possible with ITS. Instead, $F(x)$, or, at least, its logarithm, can still be evaluated reliably far beyond nine sigma, so it is not responsible for the failure of ITS, which is on $q(p)$ alone.

\subsection{Rejection sampling}

The safety of ITS depends on the precision at which quantiles can be assessed. If more refined implementations of $q(p)$ become available, the safety of ITS can be improved. Other sampling techniques that depend on different distribution descriptors might have a qualitatively distinct breakdown behavior. Rejection sampling (RS) is an alternative simulation method, but it is less general than ITS and more tailored down to specific distributions.

One Markov Chain-Monte Carlo approach related to RS is called hit-or-miss. For simulating $f_\interv$, samples can be drawn from $f(x)$ until they lie within $\interv$. The acceptance rate is $P(\interv)$. In independent sampling, it takes on average $1/P(\interv)$ samples from $f$ to generate a single sample from $f_\interv$, the number of samples is geometric-distributed. This technique is thus inefficient under extreme truncation, because such a condition makes $P(\interv)\approx 0$.

In RS, the generic distribution $f(x)$ is the target from which one wishes to draw samples, but a so-called envelope distribution must be found, denoted by $g(x)$, which must be (i) easier to simulate from and (ii) such that
$$f(x) \leq k \, g(x)\,,\quad x\in\support \,,$$
for some known constant $k \geq 1$. Here, $g(x)$ must be defined with respect to the same measure as $f(x)$, that is, the former must be either discrete or continuous in the same fashion as the latter. In general it will hold $k > 1$, while $k=1$ holds when $g(x)=f(x)$, but $f(x)$ cannot be sampled directly by assumption. In RS terms, hit-or-miss targets $f_\interv$ via an envelope $f$, which implies $k=1/P(\interv)$. As a matter of efficiency, $k$ should be small, as it takes $k$ samples from $g(x)$ to generate just one sample from $f(x)$, on average. The RS requires to draw a sample, say $x$, from $g(x)$ and accept it with probability $p(x)$, defined as
$$p(x)=\frac{f(x)}{k \, g(x)} \in \,[0,1] \,.$$
If the sample is rejected, a new one is drawn, and so on, until acceptance. The RS relies on the fact that, if $X \sim g(\cdot)$ and $U \sim \mathrm{uniform}[0,1]$ are independent, then the distribution of $X$ conditional to $U \leq p(X)$ is $f(x)$.

ITS relies on $q(p)$ as a distribution descriptor, RS on $f(x)$. Under truncation, RS relies on $\trunc{f}(x)$ and thus also on $F(x)$. Even so, RS must have a different breakdown behavior, depending on how the distribution descriptors are implemented. One can evaluate $q(p)$ by solving $F(x)=p$ for $x$; even so, some iterative procedure must be involved, which can be either cumbersome or unreliable in tails \cite{Ridout_2008}. As to RS, a major drawback is that finding an envelope is rarely trivial, at least when efficiency is a concern. Typical envelopes are often efficient around the mode $m$, in the sense that $p(m)\approx 1$, but also heavy tailed and thus potentially inefficient if reused via hit-or-miss under extreme truncation. In such a case, an entirely different envelope must be devised. For instance, the \texttt{R} package \texttt{tmvtnorm} \cite{tmvtnorm} features a hit-or-miss default sampler, which repeatedly draws samples from the non-truncated distribution until they fall within the target interval or region, thus may perform poorly in tails. The package also offers a Gibbs sampler as a replacement for such situations, as it is known to be effective even under complicated truncation \cite{Gelfand_1992}; the problem with this approach is though that it is not exact method and yields potentially correlated samples that call for inspection.

\subsection{Devroye sampling}

RS has the capability to outlive ITS under extreme truncation, because it relies on qualitatively different distribution descriptors, $f(x)$ and $F(x)$, which are numerically more stable in standard software. Yet, RS requires a smart enveloping strategy, while ITS works automatically. There exists, though, a wide range of probability models for which a rather general envelope works, with acceptance rate lower-bounded and greater than 20\%. Some of the most famous distributions are log-concave \cite{Evans_2000}. To begin with, all Tweedie distributions, like binomial and normal, are log-concave. The gamma distribution is especially tricky \cite{Philippe1997} as it is not log-concave when the shape parameter is less than unity, but it can be simulated based on the exponential power distribution \cite[Example 2.1]{Devroye_1986}. Devroye provides two lists with some log-concave distributions for the continuous case \cite{Devroye_1986} and the discrete one \cite{Devroye_1987}. Also, Devroye addresses the ideal setting, where ``real-value arithmetic can be performed with infinite accuracy" \cite{Devroye2012} but we show that, even in finite-precision calculus, DS is superior to ITS wherein applicable.

Formally, a probability mass or density function $f(x)$ is said to be log-concave if the function $\ell(x) = \log f(x)$ is concave. In the continuous case, the second derivative of $\ell(x)$ can be checked. Cusps like in the Laplace case are an exception. In the discrete case, it must hold
$$2\ell(x) \geq \ell(x-1)+\ell(x+1) \,,\quad\text{for all}\quad x \in \support\,.$$
Now, let $m$ be the mode of $f(x)$. This is a well-defined distribution summary in the case of log-concavity. Rejection sampling à-la Devroye (DS), in the continuous case \cite{Devroye_2006}, relies on the following notable inequality:
\begin{equation}\label{eq:logconcenvel}
f(x) \leq f(m) \min\left[1,\exp\left\{1-f(m)|x-m|\right\}\right] \,,
\end{equation}
or, for the unconventionally standardized variable $Y=(X-m)\,f(m)$,
$$Y \sim f(y) \leq \min\{1,\exp(1-|y|)\} \,.$$
This bound can be read in terms of a uniform-exponential mixture that can serve as an envelope for all log-concave densities, with acceptance rate exactly equal to 25\%. The simplicity of the algorithm deserves noticing \cite{Devroye_1996}. A concentration inequality is also implied, where (i) the mode turns out to be a meaningful central tendency index and (ii) the tails are necessarily sub-exponential \cite{Saumard2014}. In the discrete case, analogously, an enveloping uniform-geometric mixture can be found, with acceptance rate $\{4+f(m)\}^{-1}$ and thus between 20\% and 25\%, the two bounds being attained only in the limit as $f(m)$ approaches either $1$ or $0$, respectively.

An appealing aspect of log-concavity is that it is preserved under truncation \cite{Evans_2000}, so one can sample log-concave distributions via DS in both base and truncated versions. As a remark, even in the latter case, the acceptance rate is still at least 20\%. The mode of the truncated distribution is obtained by projecting $m$ onto $\interv$, namely, $m \leftarrow \max\{a,m\}$ \cite{Maatouk_2016}.

Algorithms \ref{devroyecont} and \ref{devroyedisc} are for Devroye's sampler in the continuous case and the discrete one, respectively, for the generally useful case $\interv=\support\,\cap\,]a,b]$. The two algorithms are collected here for completeness and adapted for coherence with our notation. One can refer to Devroye's work, in the specific, Rejection method for log-concave densities. Exponential version \cite{Devroye_1986}, and the second unnumbered algorithm from his work on the discrete case \cite{Devroye_1987}. The most relevant shift in truncation is that $F(x)$ is involved in the normalization constant to the truncated distribution and acts as a potential source of numerical nuances. In the limit as $x\to\infty$ diverges, it holds $f(x),1-F(x)\to 0$, so in \texttt{R} software the logarithmic versions of these functions are often provided, which are numerically more reliable. The relative (instead of absolute) evaluation error of $f(x)$ and $1-F(x)$ is bounded, by bounding the absolute error on their logarithms, which ensures a reasonable number of significant digits. As a general recommendation, such logarithmic version of $f(x)$ and $1-F(x)$ should be implemented meaningfully in this respect.

\begin{algorithm}\label{devroyecont}
	\caption{DS for a truncated continuous distribution.}
	\KwData{$m$, $a$, $b$, $f_{\cdot,\cdot}(\cdot)$.}
	$m \leftarrow \max\left\{a, \min\left(b, m\right)\right\}$\;
	$c \leftarrow \trunc{f}(m)$\;
	\Repeat{$\exp(Z) \leq \trunc{f}(X)$}{
		generate $U\sim \mathrm{uniform}[0,2],V \sim \mathrm{exponential}(1)$\;
		\eIf{$U \leq 1$}{
			$X \leftarrow U$\;
			$Z \leftarrow -E$\;
		}{
			generate $E^* \sim \mathrm{exponential}(1)$\;
			$X \leftarrow 1 + E^*$\;
			$Z \leftarrow -E -E^*$\;
		}
	generate $S \sim \mathrm{uniform}[-1,+1]$\;
	$S \leftarrow \mathrm{sign}(S)$\;
	$X \leftarrow m + S \, X/c$\;
	}
	\Return $X$\;
\end{algorithm}

\begin{algorithm}\label{devroyedisc}
	\caption{DS for a truncated discrete distribution.}
	\KwData{$m$, $a$, $b$, $f_{\cdot,\cdot}(\cdot)$.}
	$a \leftarrow \lfloor a \rfloor$\;
	$b \leftarrow \lfloor b \rfloor$\;
	$m \leftarrow \max\left\{a+1, \min\left(b, m\right)\right\}$\;
	$c \leftarrow \trunc{f}(m)$\;
	$w \leftarrow 1 + c/2$\;
	\Repeat{$W\min\left\{1, \exp(w - c \, X)\right\} \leq \trunc{f}(m+X) / c$}{
		generate $U,W \sim \mathrm{uniform}[0,1]$\;
		\eIf{$U \leq w/(1+w)$}{
			generate $V \sim \mathrm{uniform}[0,1]$\;
			$X \leftarrow w \, V / c$\;
		}{
			generate $E \sim \mathrm{exponential}(1)$\;
			$X \leftarrow (w + E)/c$\;
		}
		generate $S \sim \mathrm{uniform}[-1,+1]$\;
		$S \leftarrow \mathrm{sign}(S)$\;
		$X \leftarrow S \, \mathrm{round}(X)$\;
	}
	$X \leftarrow m + X$\;
	\Return $X$\;
\end{algorithm}

When a sampler fails, variates can be imputed reasonably in the light of probability calculus. For log-concave distributions, an excess sequence $Y_a$ can be defined as $Y_a \sim X - a | X > a$, whose both mean and variance are non-increasing with $a$ \cite{Bagnoli,Burdett}. Imputing a truncated random variable as infinite is not a sound choice, since $Y_a$ is bounded in probability, but this is what happens with the \texttt{R} package \texttt{truncdist}. This pastiche only occurs because the package works as a wrapper and uses probability-related functions provided by other packages in turn, so it does not address log-concave distributions differently from any other. A more meaningful imputation would be the mode $m$, after the concentration inequality in \eqref{eq:logconcenvel}. This latter kind of imputation is adopted, for instance, by the \texttt{R} package \texttt{truncnorm}. As an alternative, stochastic, imputation, one may consider exponential approximations at least for normal \cite{Geweke_1986} and perhaps other sub-exponential tails.

\subsection{Means of comparison}

Typically, the functions $f(x)$, $F(x)$ and $q(p)$ will depend on some parameter vector $\theta\in\mathbb{R}^d$. For instance, in the binomial case, $\theta=(n,p)^\top$, where $n$ is the size and $p$ is the event probability. We should thus write $f(x;\theta)$, $F(x;\theta)$, $q(x;\theta)$ and $m(\theta)$ but, for the sake of readability, we refrain from doing so. Still, all considerations should be considered dependent on the parameter $\theta$.

As per \eqref{eq:logconcenvel}, it is useful to think of a base log-concave distribution in simpler terms, based on a central tendency index $\mu=\mu(\theta)$ and a dispersion parameter $\sigma=\sigma(\theta)$, defined as
$$\mu=m \,,\quad \sigma = 1/f(m) \,.$$
In the normal-distributed case and other location and scale families, it is natural to define $\sigma$ as the base standard deviation, thus only \textit{proportional} to $1/f(m)$. This choice is meaningful also for other distributions that can be approximated to the normal, like binomial and Poisson ones.

With some distributions, and depending on their implementations, the breakdown of $q(p)$ precedes the one of $F(x)$, which in turn precedes the breakdown of $f(x)$. Call $\overline{a}=q(\overline{p})$ the maximum quantile that can be evaluated, while $F(x)$ is unreliable for $x>\overline{a}'$, and $f(x)$ for $x>\overline{a}''$. One can expect $\overline{a}'<\overline{a}''$ when $F(x)$ is implemented based on summation or numeric integration of $f(x)$. When $q(p)$ is implemented by solving $F(x) = p$ for $x$, one can expect $\overline{a}<\overline{a}'$.

All the distributions considered here share there being a maximum quantile or critical value, here denoted by $\overline{a}$, that can be computed reliably. More extreme quantiles are thence imputed equal to $\sup\support$ in some \texttt{R} packages. This behavior implies that simulating $\trunc{f}$ is not possible when $a>\overline{a}$. With ITS, the safety ratio $\eta$ is defined here as
$$\eta=\frac{\overline{a}-\mu}{\sigma}\,,$$
which can be read as the maximum sigma of events that can be simulated reliably. If $a<\overline{a}$, a fraction $1-\trunc{F}(\overline{a})$ of samples from $\trunc{f}(x)$ will be corrupted. When simulating $n$ variates, the probability of stepping into corrupted samples is $1-\{1-\trunc{F}(\overline{a})\}^n$. Sample corruption may thus occur even with non-extreme truncation, if $n$ is sufficiently large; by converse, if $n$ is small, issues will arise mostly only when $a\approx \overline{a}$, due to light tails. As a remark, $\eta$ depends on $\theta$, if the model is parameterized.

When generating $n$ samples of $\trunc{f}(x)$ with DS, the failure is either sure or impossible, as ruled by $\overline{a}'$, and essentially independent of $n$. With RS, we define the safety ratio $\eta'$ as
$$\eta' = \frac{\overline{a}'-\mu}{\sigma} \,,$$
without risk of overstating the algorithm's safety. We argue that DS allows for $\eta'>\eta$, that is, additional safety with respect to ITS.

There are at least two aspects in the definition of $\eta$ that are worth discussing. Firstly, it is necessary to test for the distribution of generated samples to assess its trustworthiness, but testing is as tricky as random number generation itself, because both tasks are based on some distribution descriptor that is potentially troubled. Secondly, the standardization of $\overline{a}$ into $\eta$ is meaningful at least in the normal case, since the evaluation of $f(x)$, $F(x)$ and $q(p)$ involves an internal standardization. The same definition still makes sense at least for some other distributions that can be approximated to the normal and that are treated this way in software, like Poisson and binomial.

Table \ref{tab:distros} summarizes our definition of $\mu$ and $\sigma$ along with the choice of \texttt{R} implementation of all the distributions we checked. In general, the package \texttt{Rmpfr} \cite{Rmpfr} was preferred over \texttt{stats}\nocite{R}. Both lacked an implementation for the inverse Gaussian, which we retrieved from the \texttt{R} package \texttt{actuar} \cite{actuar}.

\begin{table}
	\caption{\label{tab:distros} Distributions considered in simulations.}
	\begin{center}
		\resizebox{\textwidth}{!}{
			\begin{tabular}{lc|cc | lll}
				distribution & parameters & $\mu$ & $\sigma$ & $f(x)$ & $F(x)$ & $q(p)$ \\
				\hline
				binomial & $n,p$ & $np$ & $\sqrt{np(1-p)}$ & \texttt{Rmpfr} & \texttt{stats} & \texttt{stats} \\
				Poisson & $\lambda$ & $\lambda$ & $\sqrt{\lambda}$ & \texttt{Rmpfr} & \texttt{stats} & \texttt{stats} \\
				negative binomial & $n,p$ & $\frac{np}{1-p}$ & $\frac{\sqrt{np}}{1-p}$ & \texttt{Rmpfr} & \texttt{stats} & \texttt{stats} \\
				geometric & $p$ & $0$ & $\frac{\sqrt{1-p}}{p}$ & \texttt{stats} & \texttt{stats} & \texttt{stats} \\
				normal & $\mu,\sigma$ & $\mu$ & $\sigma$ & \texttt{Rmpfr} & \texttt{Rmpfr} & 
				\texttt{stats} \\
				gamma & $\alpha,\lambda$ & $\frac{\alpha}{\lambda}$ & $\frac{\sqrt{\alpha}}{\lambda}$ & \texttt{Rmpfr} & \texttt{stats} & \texttt{stats} \\
				inverse Gaussian & $\mu,\lambda$ & $\mu$ & $\sqrt{\frac{\mu^3}{\lambda}}$ & \texttt{actuar} & \texttt{actuar} & \texttt{actuar}
		\end{tabular}}
	\end{center}
\end{table}

\section{Software}

The existing \texttt{R} package of interest is \texttt{truncdist}, which allows to simulate truncated distributions based on ITS, in the case when the function $q(p)$ can be evaluated feasibly \cite{truncdist}. In the event of extreme truncation or other situations making $q(p)$ hard to evaluate, the \texttt{R} package \texttt{truncnorm} \cite{truncnorm} replaces ITS with RS in simulating truncated normal distributions. Many other packages will address truncation via specifically tailored RS, but this approach is not automatic enough nor of interest here. The following packages are instead based on DS.
\begin{itemize}
	\item \texttt{UnivRNG} \cite{UnivRNG} for Zipf's zeta distribution,
	\item \texttt{extraDistr} \cite{extraDistr} for Dirichlet distribution,
	\item \texttt{actuar} \cite{actuar} for inverse Gaussian distribution,
	\item \texttt{heavy} \cite{heavy} for multivariate normal and $t$ distributions.
\end{itemize}
We do not address the multivariate case here, as it likely deserves a more targeted set of strategies, the task being cursed by dimensionality \cite{Cadez2002,Maatouk_2016}. To our knowledge, the above \texttt{R} packages are the only ones available on the Comprehensive \texttt{R} Archive Network (CRAN) that simulate truncated distributions via DS. Even so, each of these packages addresses only one specific model. It seems thus that DS serves mostly as a placeholder in the absence of a more targeted sampler. Also, these package lack the purpose of addressing a wide variety of truncated distributions by leveraging on the scalability of DS.

We propose a new \texttt{R} package that we call \texttt{truncLCdist} as a hommage to \texttt{truncdist}'s aspiration to simply work out of the box. Our package is based entirely on DS and can address all log-concave distributions, provided a definition of $f(x)$, $F(x)$ and $m$, consistently parameterized. An implementation of $\log f(x)$ must be provided: this can be a foo, which firstly evaluates $f(x)$ and then computes its logarithm or, as recommended, a direct evaluation of $\log f(x)$ that ensures a reasonably high number of significant digits \cite{Loader2002}. The two steps evaluation involves firstly computing $f(x)$ and then transforming it via logarithm and may be unreliable at times: the two steps are optimized separately, without foreseeing each other, thus the overall evaluation of $\log f(x)$ is less accurate.

Tweedie distributions are particularly relevant for regression analyses and are included in our package as a default. The user can add custom support for other log-concave distributions just by linking the required functions. Such linkage must be \textit{explicitly} set, via a function that wraps functionalities borrowed from the \texttt{futile.options} package \cite{futile}. The linkage also allows to check for issues with specific functions, by handily replacing their implementations. This cannot be done as easily with \texttt{truncdist}, where the linkage is \textit{implicit}, that is, the package retrieves the required functions from the global environment and loaded packages based on names. One should first understand the way in which function masking and conflicts between packages are addressed in \texttt{R}, before being able to feed an alternative distribution descriptor to \texttt{truncdist}.

\section{Simulations}\label{sec:simulations}

Here we enumerate few simulation studies that can be repeated with our software in order to probe into the behavior of ITS and DS in practical random number generation.

\subsection{Normal}

As a starter, we address a simple case with a normal distribution, with mean $\mu$ and variance $\sigma^2$. One can consider the standard normal, with mean $\mu=0$ and variance $\sigma^2=1$, without loss of generality.

For the sake of completeness, $F(x)$ translates well under rescaling since it only depends on the standardized input $z=(x-\mu)/\sigma$, while $f(x;\mu,\sigma)=f(z;0,1)/\sigma$ is also divided by $\sigma$ and may thus be troubled in the limit as $\sigma\to 0$ or $\sigma\to\infty$. We assume here that this issue does not occur here and that $\sigma$ takes non-extreme values, for the sake of simplicity. We defend ourselves with this remark as even our proposal may fail arbitrarily subtle tests, but our approach is still motivated by the usefulness of truncated distributions that are extreme in terms of $\interv$.

We probe some integer values of $a$ and look for its largest value $\overline{a}$ for which $q(p)$ is safe to evaluate. Here, it holds $\eta=\overline{a}$, since we probe the standard normal only. Then, ITS has a safety margin $\eta \leq 7$ while DS has $\eta' \leq 38$. These results are conditional to using the implementations of $f(x)$, $F(x)$ and $q(p)$ available in \texttt{Rmpfr}, but they do not change when falling back to the original implementations provided in \texttt{stats}.

These results can partly explain the safety margins for other distributions, as \texttt{R} basic libraries often utilize normal approximations. In particular, base \texttt{R} implementations of binomial and Poisson $q(p)$ involve a so-called Edgeworth approximation, which is essentially an enhancement modification to approximate normal quantiles for the case of skewed distributions. We address these approximately normal distributions in the following subsections.

\subsection{Binomial }

\begin{figure}
	\caption{\label{fig:binomial} Safety in the binomial$(n,p)$ case. The x-axis is on a logarithmic scale, the y-axis is on a logit scale.}
	\includegraphics[width=\linewidth]{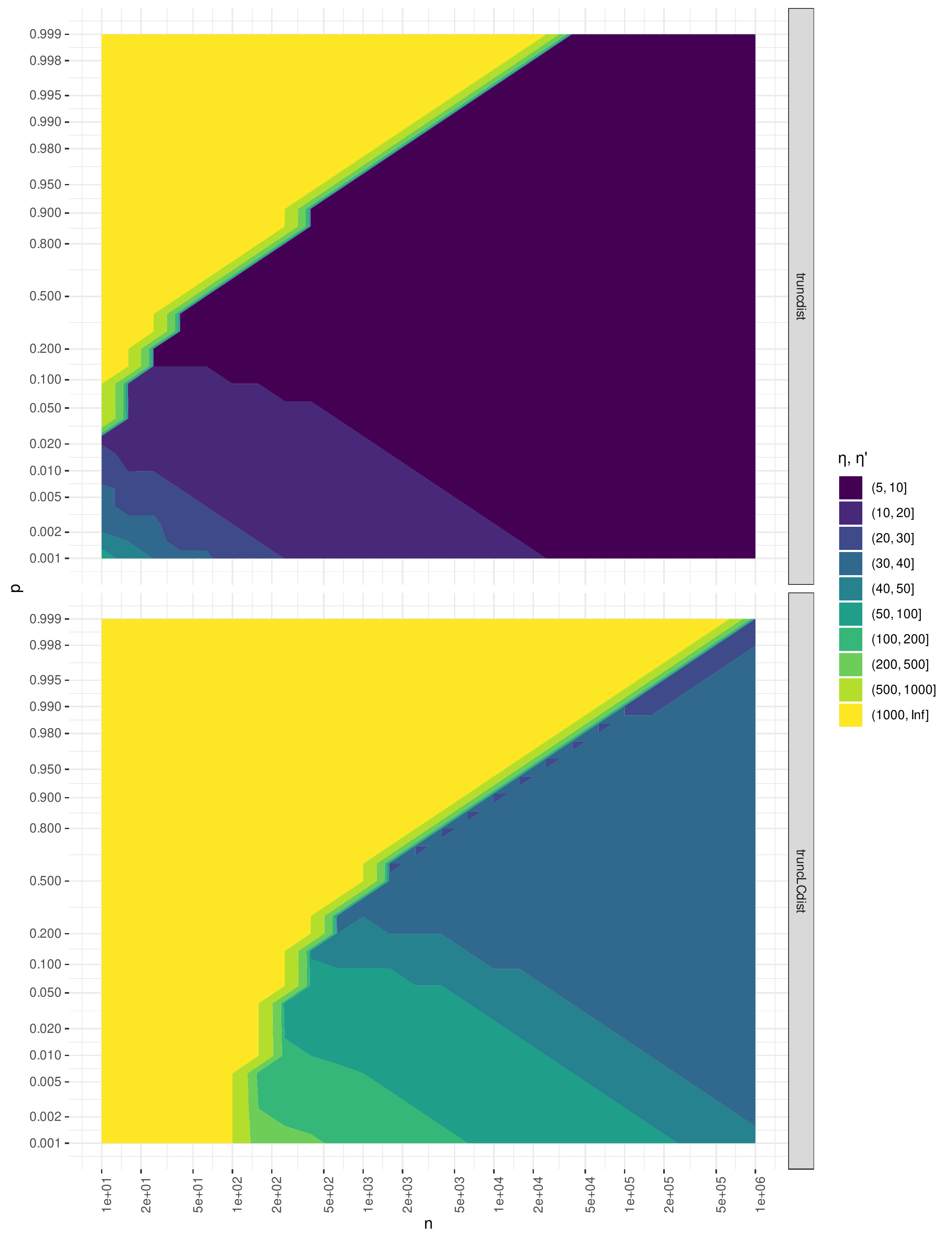}
\end{figure}

We consider now the binomial case. We ran simulations under some configurations of its parameters, namely, size $n$ and probability $p$, with mean $np$ and variance $np(1-p)$. In Figure \ref{fig:binomial}, the values of $\eta$ and $\eta'$ were obtained at few selected locations and then interpolated for added readability in maps.

A troubled cone-shaped region is visible in the middle of these plots, with $p=1/2$ as the most difficult iso-$n$ configuration, and issues spreading to more values of $p$ as $n$ diverges. It seems that, in the limit, \texttt{truncdist} has safety margin limited to $\eta \leq 10$, likely due to internal approximations of $q(p)$ to the normal.

The cone visible in plots is mostly a reflection of a condition on $n$ and $p$ that triggers such approximation, so it is partly a side-effect of a specific implementation. In this sense, we trust a sampler as far as it can simulate the approximate, normal, distribution, so there is room for more relevant tests than the ones we present here, which though require a different and more precise binomial implementation.

The proposed alternative \texttt{truncLCdist} package not only provides better support for small values of $n$, but it does achieve a higher safety margin as $n$ diverges, where $\eta' \leq 40$. This result, though close to the one for normal, is instead related to the beta distribution, due to the way in which binomial $F(x)$ is approximated in \texttt{stats}. This aspects makes the binomial example not redundant after results related to the normal case.

\subsection{Poisson}

\begin{figure}
	\caption{\label{fig:poisson} Safety in the Poisson$(\lambda)$ case. Both x- and y- axis are on a logarithmic scale.}
	\includegraphics[width=\linewidth]{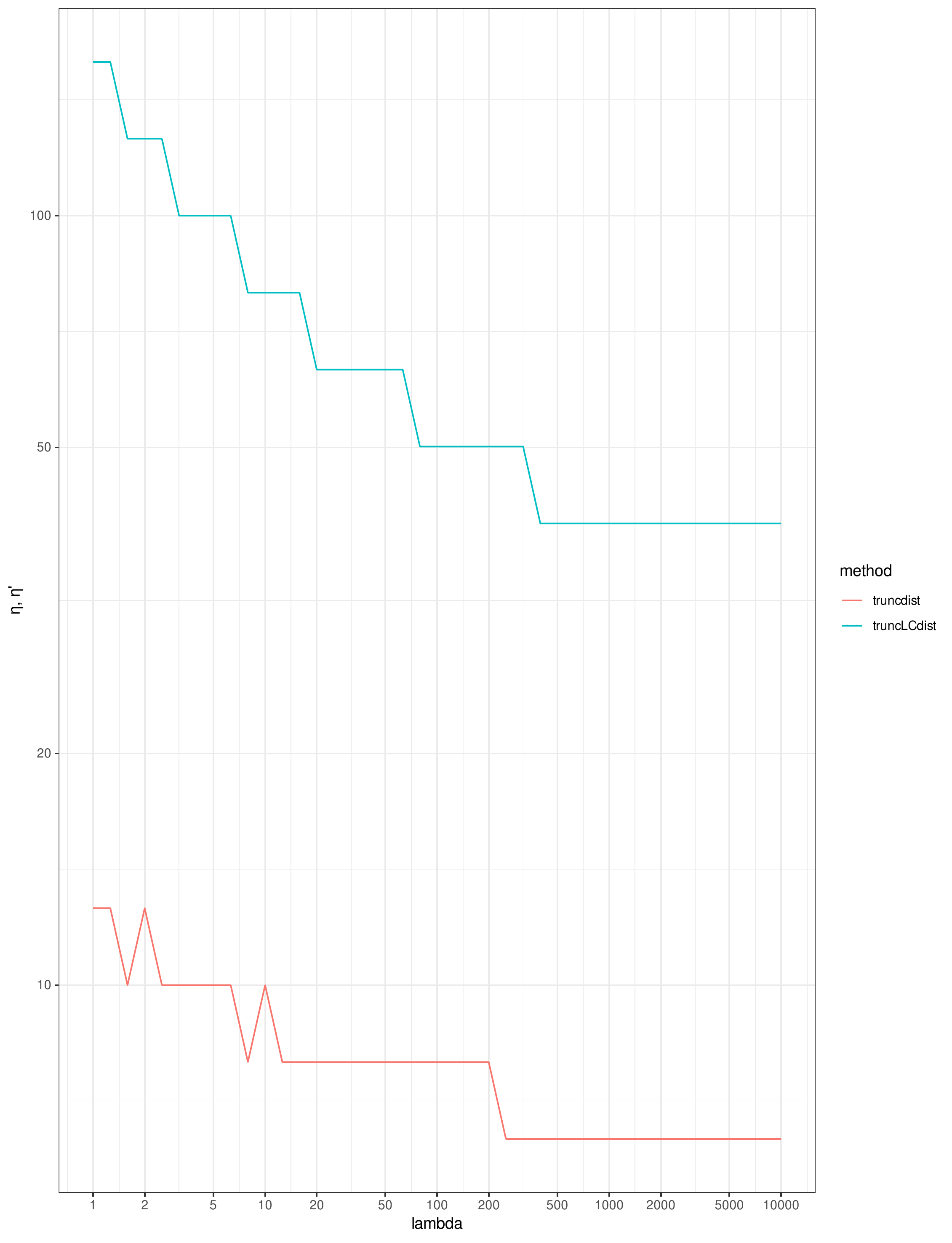}
\end{figure}

Poisson distribution is ruled only by a shape parameter $\lambda>0$ representing both the mean and the variance. In Figure \ref{fig:poisson}, the safety margin $\eta$ is visibly under ten sigma with \texttt{truncdist}. This result too might depend on normal approximations used in \texttt{stats}. As to \texttt{truncLCdist}, $\eta'$ is again close to the critical value $\eta'\approx 38$ of the normal case. This result is effectively redundant, since the Poisson distribution ships with an approximation to the normal in \texttt{R}.

\subsection{Remarks}

As a nuance, truncation cannot be arbitrarily extreme in simulations and the safety margin is typically undocumented. This issue does not arise in abstract probability calculus, but in practical finite-precision computing, which is why we had to simulate in order to make our point. Precision in such computations is limited, so $F(x)$ is imputed equal to either zero or one if $X=x$ is a many sigma event. Its inverse $q(p)$ cannot be evaluated meaningfully for all orders, relatedly. This problem must be more acute for the normal and the other light tailed distributions, as their tail probabilities decay faster, but it can be expected to affect even Cauchy and other heavy tailed distributions at some point.

These results serve as a reminder that even basic building blocks in statistics, such as normal quantiles and tail probabilities, can still be a topical issue. When implementing probability-related functions in statistical software, care must be taken for the consequences of low precision in tasks that push these mathematical objects to their computational limits. With truncated distributions, safety should not be taken for granted, the behavior of ITS should be checked, and alternative samplers should be considered.

\subsection{Geometric}

The geometric distribution is ruled by the event probability $p$, the \texttt{R} default implementation has support over all non-negative integers, and it is endowed with a property known as \textit{lack of memory}, namely,
$$X \sim \mathrm{geom}(p) \implies P(X > x + h|X > x) = P(X > h) \,,\quad x,h = 0,1,\dots \,.$$
Truncation under a geometric model just implies an affine translation of the random variable. Even this simple fact from probability calculus may not be reflected in computation due to software-related issues. The package \texttt{truncdist} is agnostic as to the probability distributions, giving no special attention to anyone in the specific, so does our \texttt{truncLCdist} package.

\begin{figure}
	\caption{\label{fig:geom} Safety in the geometric$(p)$ case. The x-axis is on a logit scale, the y-axis on a logarithmic scale.}
	\includegraphics[width=\linewidth]{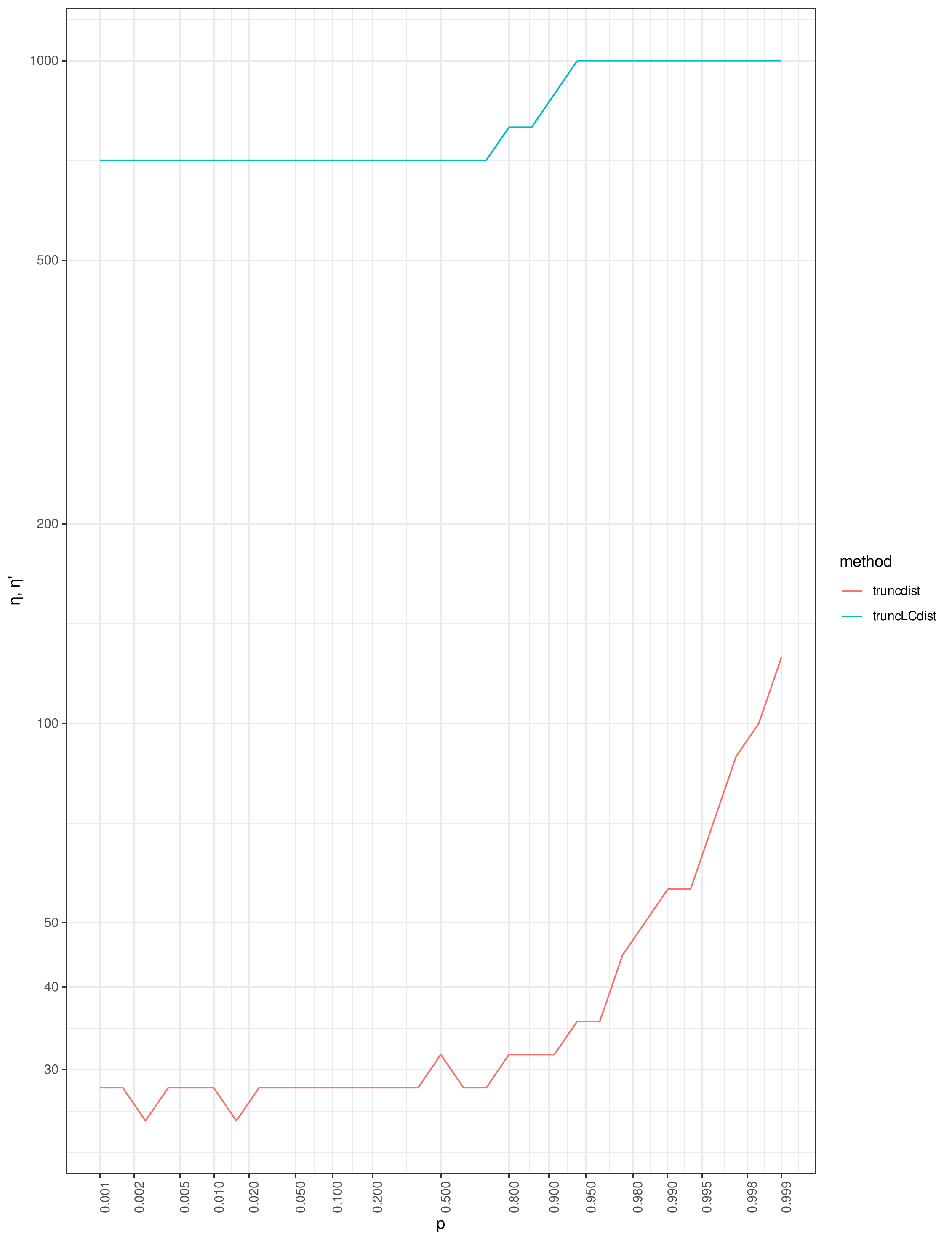}
\end{figure}

We probe some configurations of $p$ and check some values of $a$ taken from a geometric progression, until failure is detected. In Figure \ref{fig:geom}, the value of $\eta$ is reported as a function of $p$. One may first notice the asymmetry in the plot, as among opposite log-odds the positive ones are easier to deal with. It is disturbing how little the absence of memory is reflected in ITS-based simulations. While such a property should hold independently of the extremeness of truncation, it is at least reassuring that DS can reflect it to a broader extent.

\subsection{Exponential}

The exponential distribution is ruled by the rate parameter $\lambda>0$, whose reciprocal serves as the scale parameter $\sigma$, and the latter can be kept fixed in simulations with little loss of generality, in the same fashion as when keeping $\sigma$ fixed in the normal case. The exponential distribution has no memory, so it is the continuous counterpart to the geometric distribution. In formulas,
$$X \sim \mathrm{exponential}(\lambda) \implies P(X > x + h|X > x) = P(X > h) \,,\quad x,h \geq 0\,.$$

We simulated the exponential distribution via the gamma with shape parameter equal to one. Checking all integer values of $a$ from $1$ to $1000$, ITS seems safe up to $\eta \leq 31$, while DS is safe to a broader extent, as $\eta' \leq 745$.

\subsection{Exponential power distribution}

We shortly introduce the exponential power distribution (EPD), which is supported by our package only for ancillary purposes. Let the EPD be defined as
$$f(x;\beta) = \frac{1}{2\Gamma(1/\beta + 1)}\exp\left(-|x|^\beta\right) \,,\quad x \in\mathbb{R}\,.$$
When $\beta \geq 1$, the distribution is log-concave, and this is the only case relevant to us, wherein it allows to generate gamma-distributed variates, indirectly.

\subsection{Gamma}

The gamma distribution is ruled by a shape parameter $\alpha>0$ and the rate $\lambda>0$, hence it has expected value $\alpha/\lambda$ and variance $\alpha/\lambda^2$. In simulations, we probe a grid of values for $\alpha$, while we keep $\lambda=1$ fixed without loss of generality, as with the scale parameter in the normal case.

The gamma distribution is a special one among those we address in this paper and in the package, since it is a log-concave distribution only for some parameter configurations, namely, when $\alpha\geq 1$. The case $\alpha=1$ implies an exponential density. As to the case $\alpha<1$ \cite{Devroye_1986}, if $X$ is EPD with shape parameter $\beta>1$ then $|X|^\beta$ is gamma-distributed with $\alpha=1/\beta<1$. In the package we set up an exception handler that allows to override an existing definition depending on the parameter configuration.

\begin{figure}
	\caption{\label{fig:gamma} Safety in the gamma$(\alpha,\lambda=1)$ case. Both x- and y- axis are on a logarithmic scale.}
	\includegraphics[width=\linewidth]{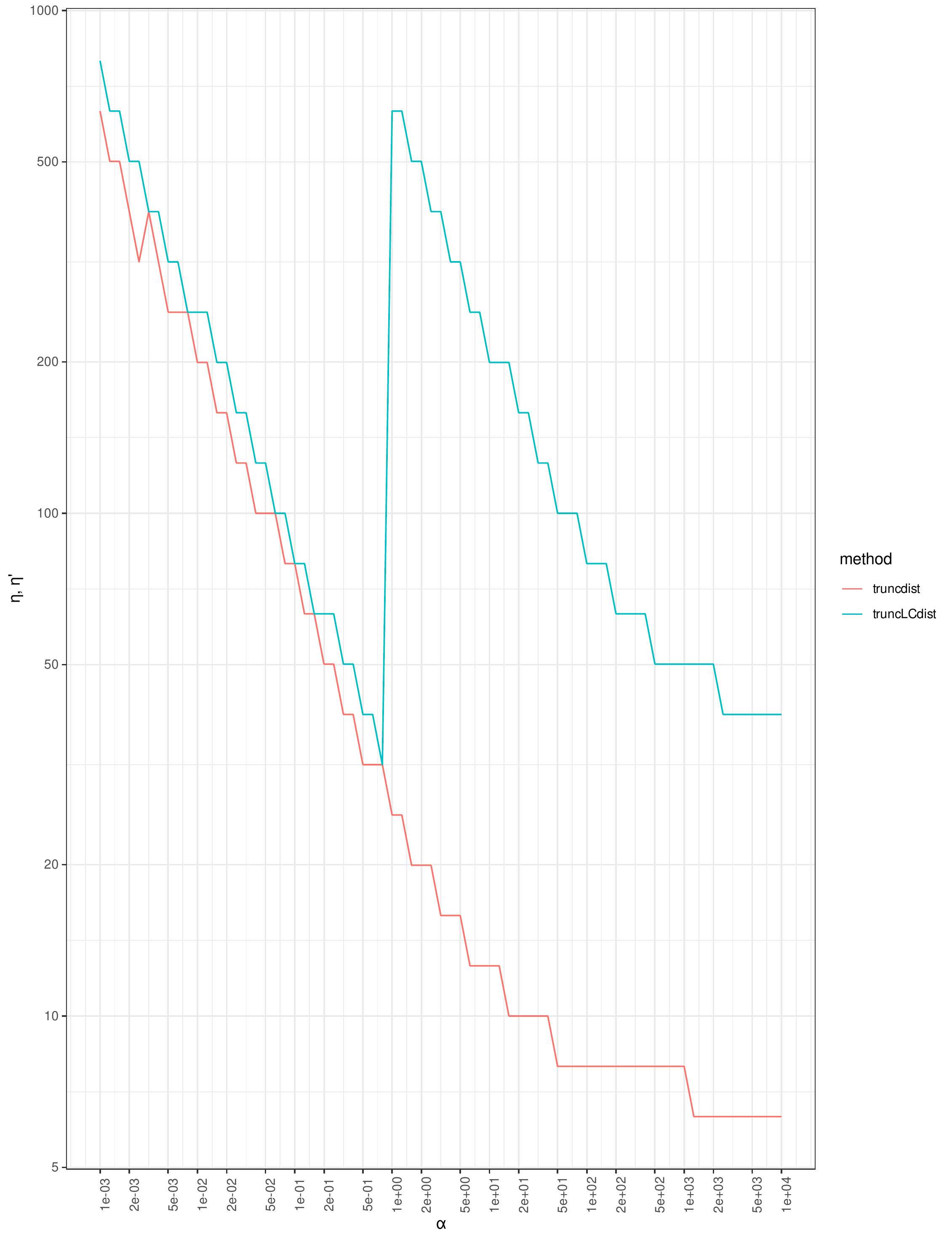}
\end{figure}

Figure \ref{fig:gamma} reports on $\eta$ and $\eta'$ for different configurations of $\alpha$. When $\alpha<1$, DS seems not to improve much on traditional ITS. Differences in safety can be appreciated, instead, when $\alpha\geq 1$. This otherwise curious shift in regime around $\alpha=1$ is only due to the switch in implementation from the indirect EPD-based simulation when $\alpha<1$ to direct DS in the case when $\alpha\geq 1$.

As a matter of style, this casewise solution may not be elegant, but the gamma case looks actually tricky in the literature, where many other samplers have been proposed to address exclusively one of the two cases. As it concerns the generality of the exception handler, this workaround may look mostly like a patch meant only for the gamma case, but we foresee it may help in implementing other distributions that are log-concave only on subspaces of $\support$ (such as the $t$-distribution) or even transformations of log-concave-distributed variables (which may not be log-concave-distributed in turn).

\subsection{Negative binomial}

The negative binomial distribution, also known as compound Poisson-gamma, can be used in modeling the length of a sequence of Bernoulli random variables, with event probability $p$, ending upon the $n$-th event. Like in the binomial case, $n$ and $p$ are shape parameters, so we probe a two-dimensional grid of configurations.

\begin{figure}
	\caption{\label{fig:nbinom} Safety in the negative binomial$(n,p)$ case. The x-axis is on a logarithmic scale, and the y-axis is on a logit scale, like in the binomial case.}
	\includegraphics[width=\linewidth]{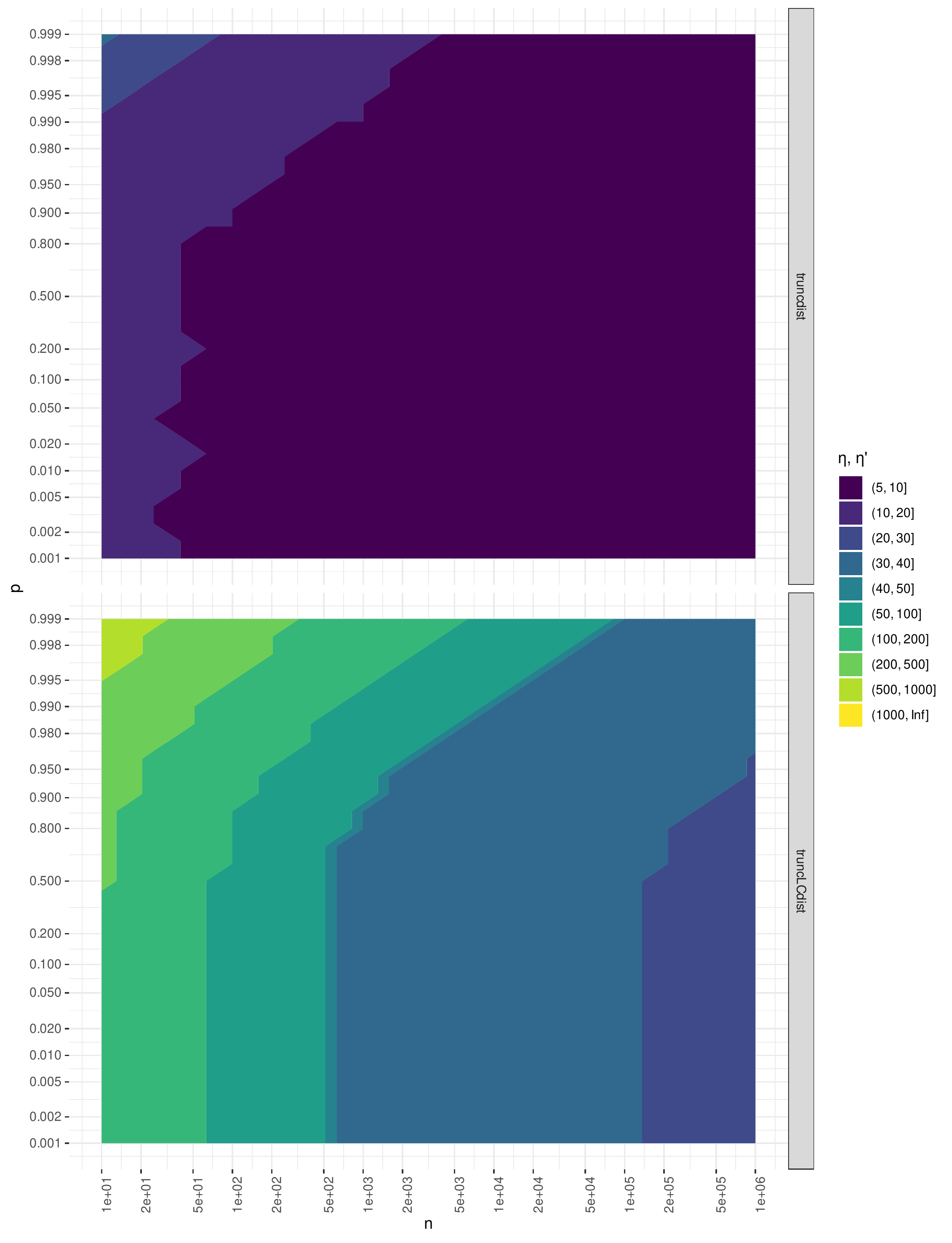}
\end{figure}

The values of $\eta$ are reported in Figure \ref{fig:nbinom}. DS looks uniformly safer than ITS, though the safety of both samplers visibly erodes as $n$ diverges. Here again $q(p)$ fails sooner than $F(x)$, hence the relation between the contour plots.

\subsection{Inverse Gaussian}

The inverse Gaussian distribution is ruled by two orthogonal shape parameters, $\mu>0$ and $\phi>0$, respectively the mean and dispersion parameters. The safety of DS looks uniformly superior to the one of ITS.

\begin{figure}
	\caption{\label{fig:ingauss} Safety in the inverse Gaussian$(\mu,\phi)$ case. Both x- and y- axis are on a logarithmic scale.}
	\includegraphics[width=\linewidth]{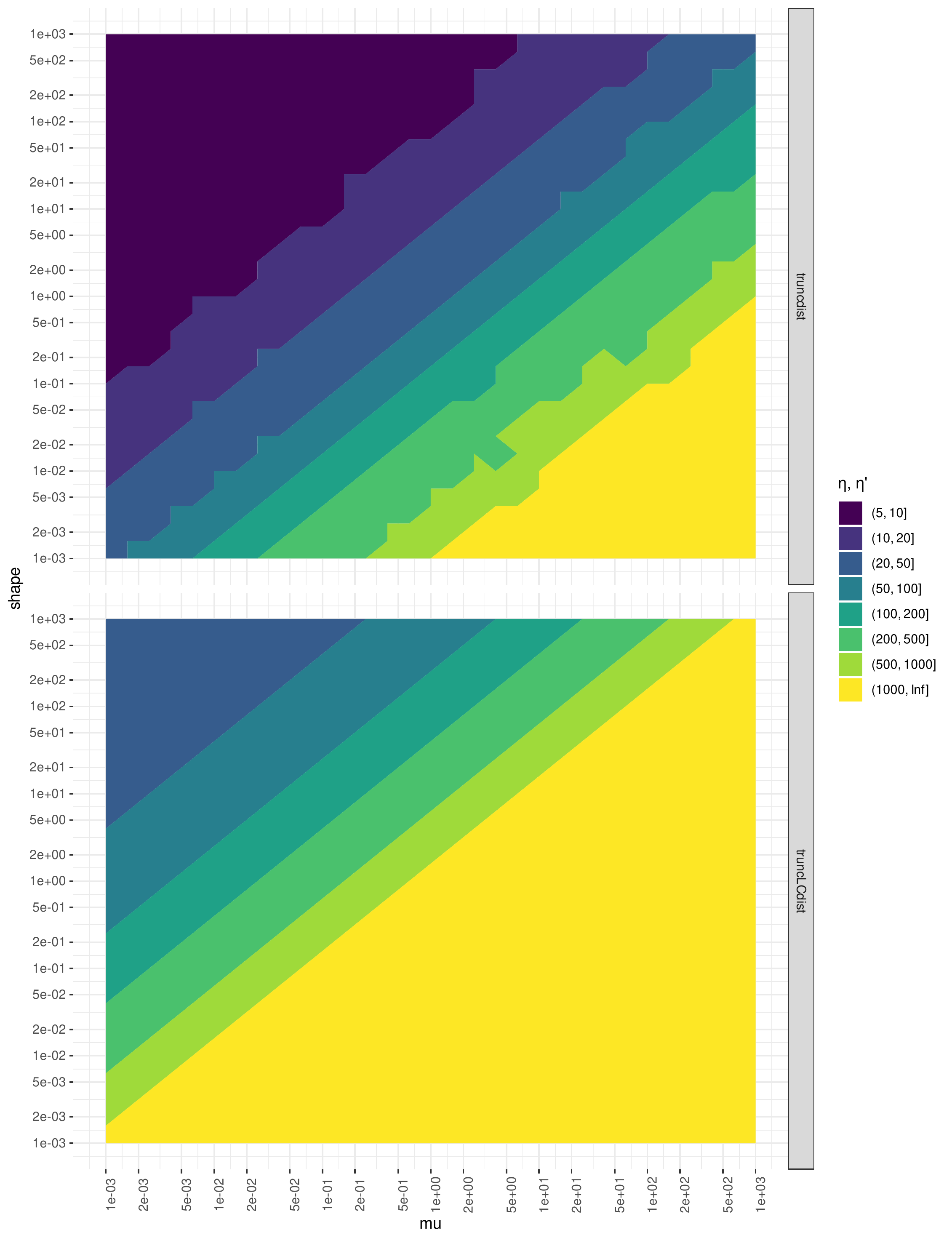}
\end{figure}

\section{Beyond diagnostics}\label{sec:diag}

In the previous section, we have investigated the safety margins of two samplers based on some proxy information, namely, diagnostics. Our endpoint with \texttt{truncdist} was the presence or absence of execution errors or infinite values. The endpoint for \texttt{truncLCdist} was the imputation rate. However, even when diagnostics raise no warning, it is still possible that samples do not reflect the sampled distribution. We now validate the diagnostics as a proxy for actual reliability of our proposed sampler. Since an exhaustive validation is not possible, we only show few examples, namely, normal, Poisson and geometric.

\subsection{Normal}

The truncated normal distribution is quite popular in statistical modeling, as outlined in the introduction. The normal distribution is also one of the few whose truncation behavior is well known. The Mills ratio recurs in the expression of cumulants \cite{Landsman}, and the excess distribution resembles the exponential under extreme truncation \cite{Geweke_1986}. The behavior of other distributions is generally less known, but this knowledge can be surrogated with Monte Carlo simulations. As a remark, this is why the reliability of samplers is so crucial.

We consider the standard normal case, without loss of generality, because the mean and standard deviation serve as location and scale parameters, respectively. We let $a$ range in the lattice $0,0.05,0.1,\dots,50$. Both the samplers we consider have safety $\overline{a}$ lying between $0$ and $50$, actually. For each value of $a$, we simulate $n=10^5$ independent samples with both samplers. Ideally, it would be best to test the distribution itself in a Kolmogorov-Smirnov fashion, but the problem is that the descriptors in use for random number generation affect both simulation and the test statistic. We use the sample mean
$$\overline{X} = \frac{1}{n}\sum_{i=1}^n X_i \,,$$
and standard deviation
$$S=\sqrt{\frac{1}{n-1}\sum_{i=1}^n (X_i-\overline{X})^2} \,,$$
based on $n$ independent variates $X_1,\dots,X_n \sim f_\interv$, to compute a $Z$-statistic defined as
$$Z = \frac{\overline{X} - \mu_a}{S/\sqrt{n}} \,\dot\sim \, \mathrm{normal}(0,1) \,.$$
Here, $\mu_a=\mathbb{E}X_i>a$ is the expected value of the truncated normal, dependent on $a$. If the sampler is reliable, then $Z$ has approximately a standard normal distribution. The aim is to test the mean of the samples, while it would not be meaningful to test the mode for instance, because this is typically the imputed value with our package.

Our test with the $Z$-statistic is meaningful only as far as $\mu_a$ can be computed reliably. The \texttt{R} package \texttt{truncnorm} \cite{truncnorm} provides a function \texttt{etruncnorm} that computes $\mu_a$ reliably even under extreme truncation. However, even this function has a rupture point, as for instance it can yield $\mu_a<a$ when $a>12000$, though this does not look like an impairing defect. The evaluation of truncated variance via the function \texttt{vtruncnorm} is instead much less reliable, as it fails for $a>9$. This is why we estimate the truncated standard deviation empirically, via $S$.

\begin{figure}
	\caption{\label{fig:normz} $Z$ statistic for the truncated normal mean, $n=100000$ samples.}
	\includegraphics[width=\linewidth]{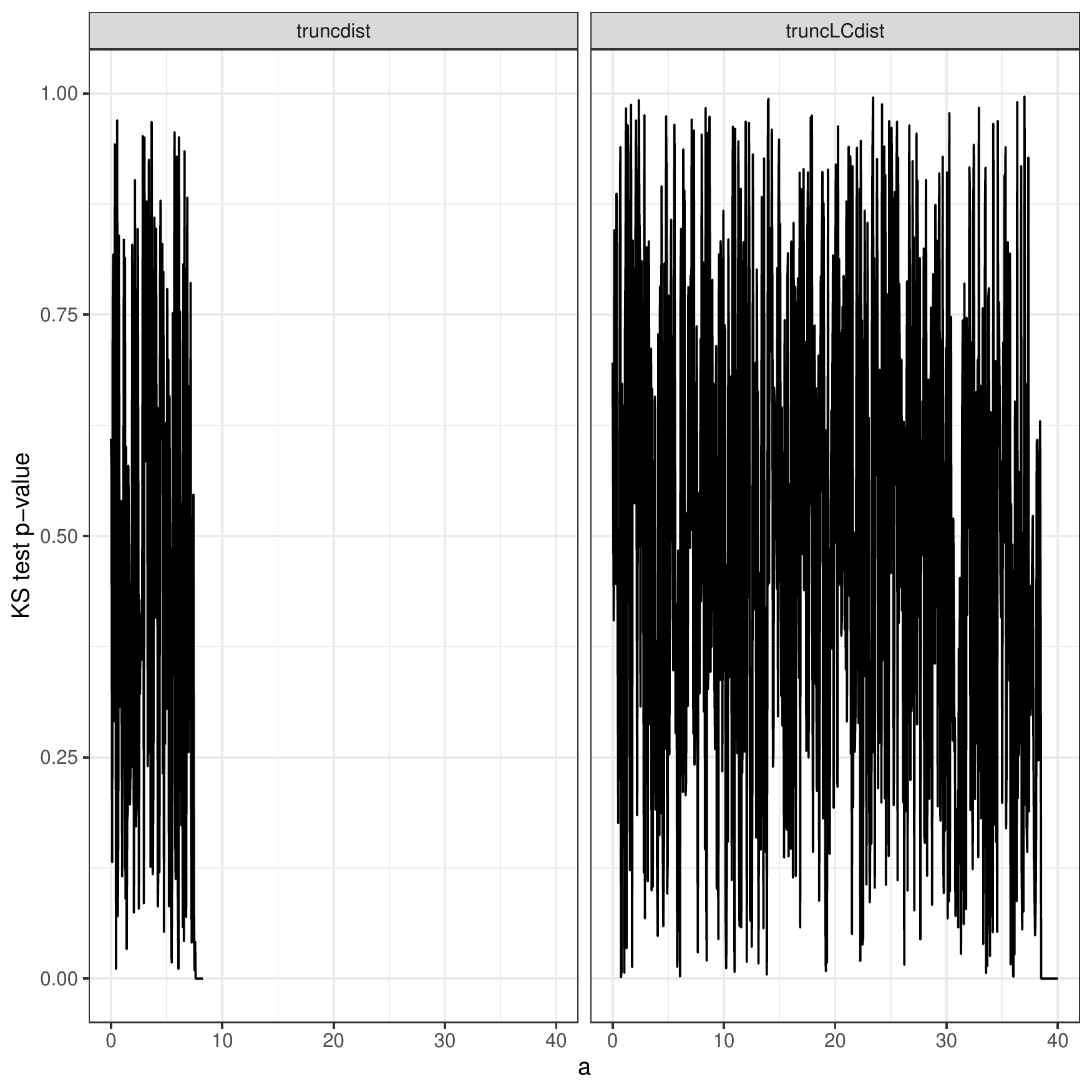}
\end{figure}

In Figure \ref{fig:normz}, the $Z$-statistic is reported along the vertical axis, and $a$ along the horizontal one. Our proposed sampler works for $a \leq \overline{a}=38.45$ while \texttt{truncdist} is still less safe as predicted. However, the point in this test is to show that our package can actually be trusted when diagnostics report no error.

\begin{figure}
	\caption{\label{fig:normecdf} Truncated normal, $a=38.45$, $10^6$ samples via \texttt{truncLCdist}.}
	\includegraphics[width=\linewidth]{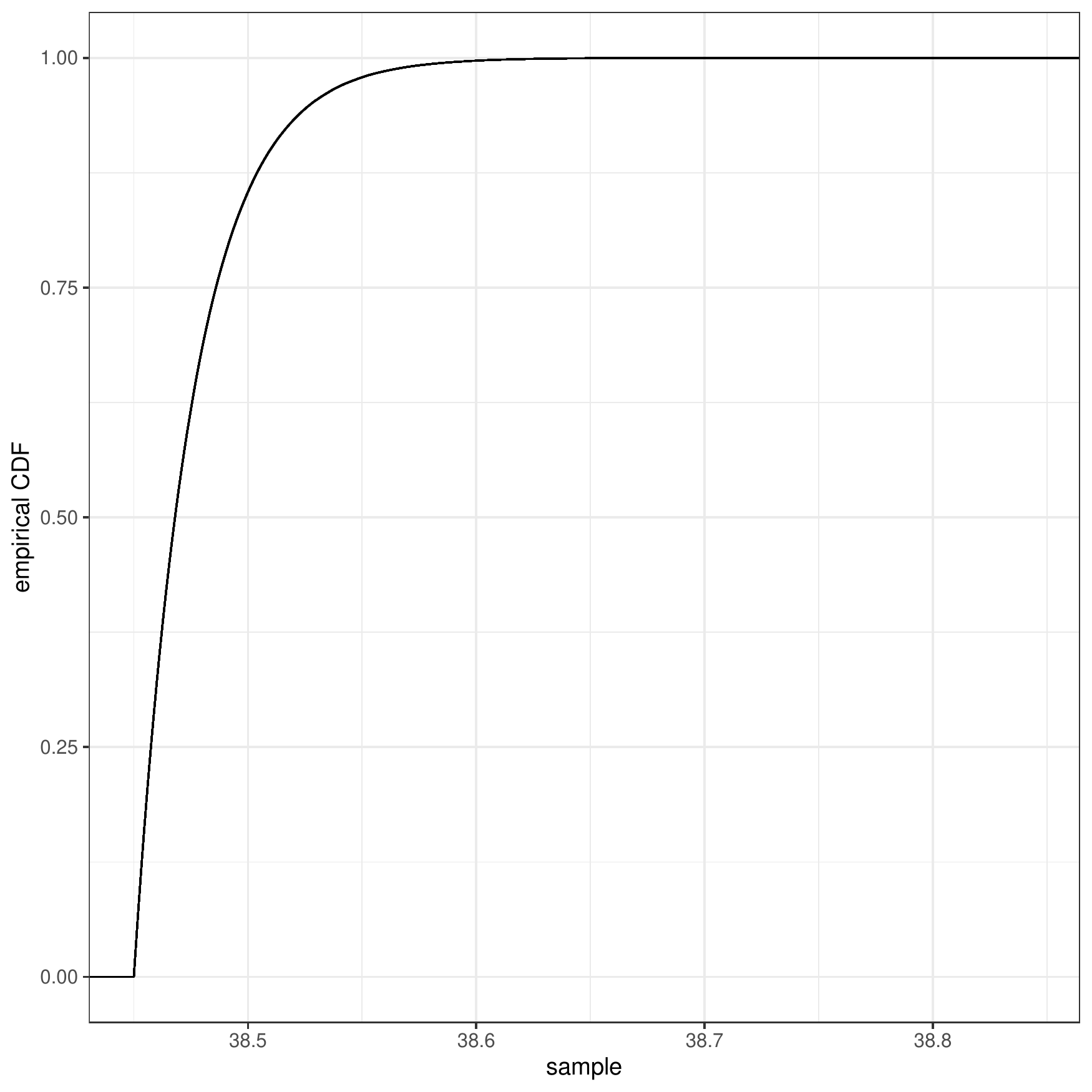}
\end{figure}

We now focus on the behavior of DS as $a=\eta'$. When truncating the standard normal over $\interv\,=\,]\eta',+\infty]$, the empirical cumulative distribution function with $n=1000000$ samples can be as in Figure \ref{fig:normecdf}. This distribution may not be assessed out of simulations as even the \texttt{R} package \texttt{truncnorm} does not represent it accurately in terms of quantiles.

\begin{figure}
	\caption{\label{fig:normqq} Truncated normal, $a=38.45$, Q-Q plot for an exponential$(a)$ null hypothesis with $10^6$ samples via \texttt{truncLCdist}.}
	\includegraphics[width=\linewidth]{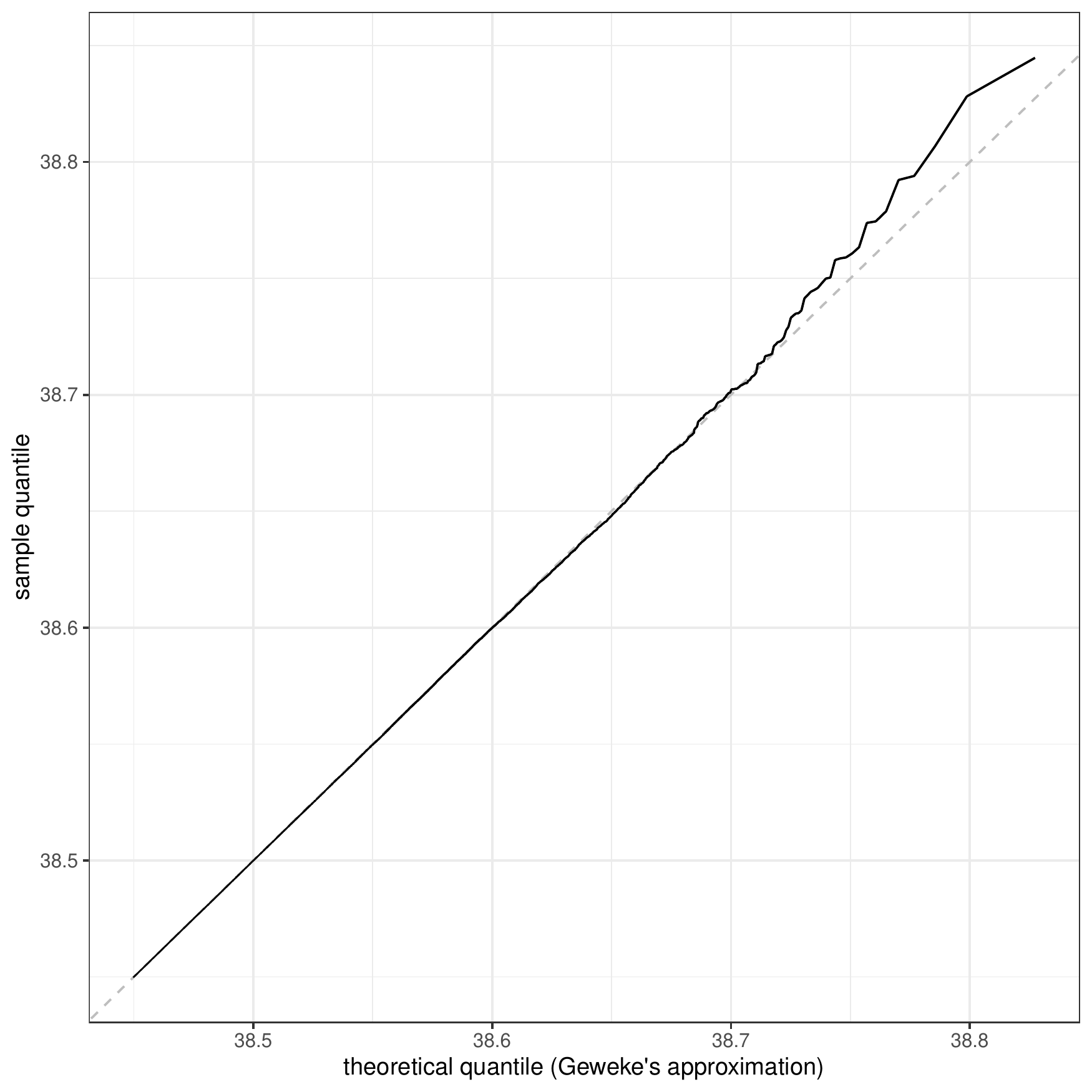}
\end{figure}

It can be shown easily that, provided $a$ is ``large", for $X \sim \mathrm{normal}(0,1)$ it holds \cite{Geweke_1986}
$$X - a|X > a \,\approx\, \mathrm{exponential}(a) \,.$$
This exponential approximation seems to hold in Figure \ref{fig:normqq}, where we report the Q-Q plot for sample quantiles versus the approximate theoretical ones.

\subsection{Poisson}

As a discrete case, we consider the Poisson distribution. In the normal case it was possible to set the scale and location parameters arbitrarily without loss off generality, while Poisson distribution is non-trivially affected by a shape parameter $\lambda>0$. We then probe a grid of values $(\lambda,z)$, and we set $a=\lambda+z\sqrt{\lambda}$. For each configuration, we compute the mean and standard deviation on $100000$ draws from $X\sim\mathrm{Poisson}(\lambda)$ truncated over $]a,+\infty]$. We use these data to test the samples for Poisson truncated mean, which is defined as follows, see for instance \cite{Landsman}.
$$\lambda(a)=\mathrm{E}(X-a|X>a) = \lambda \cdot \frac{1-F_X(a-1)}{1-F_X(a)}\,.$$

\begin{figure}
	\caption{\label{fig:poissonz} $Z$ statistic for the truncated Poisson mean, $n=100000$ samples for each configuration of $(\lambda,z)$. A conjectured asymptote at $z=38$ is also shown.}
	\includegraphics[width=\linewidth]{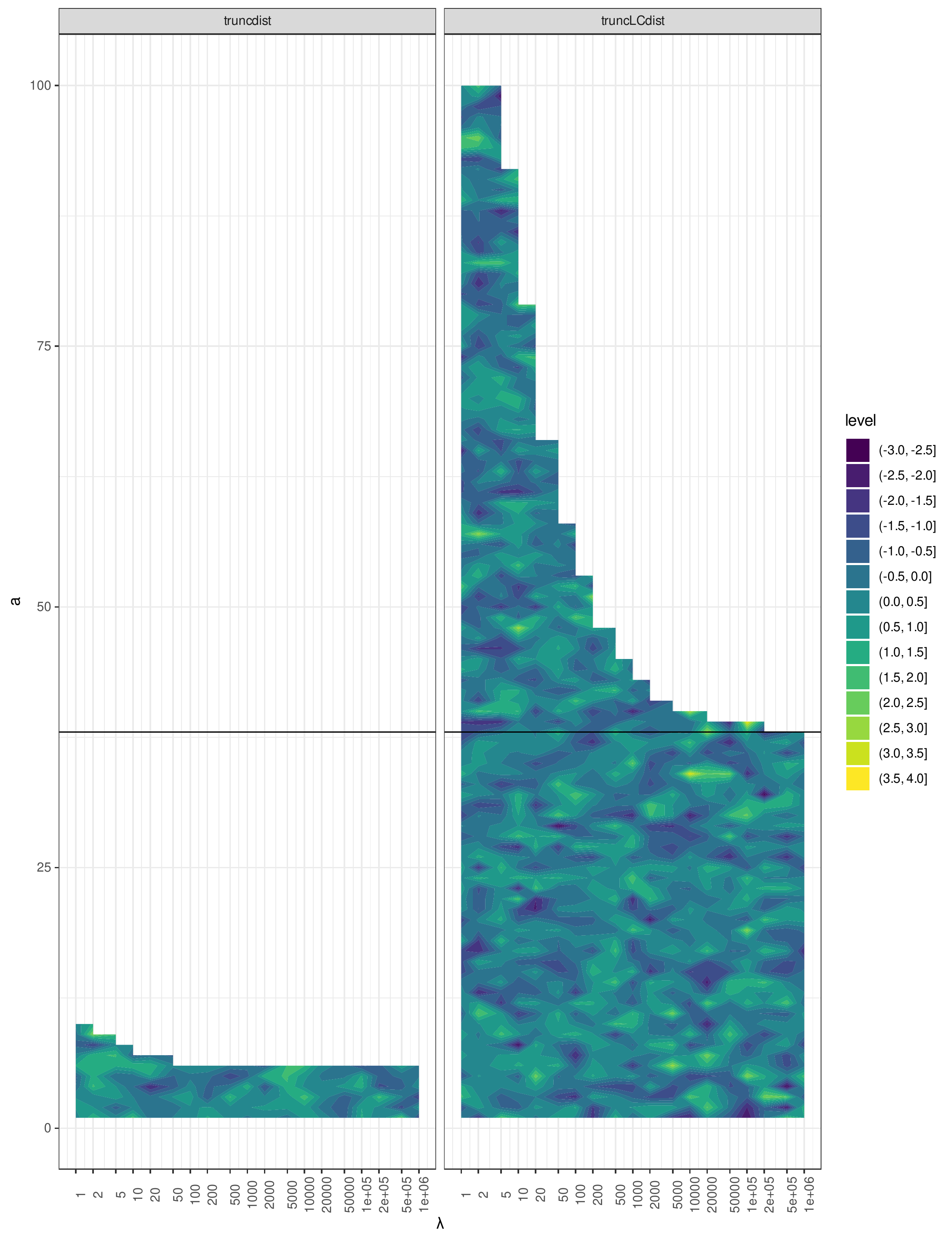}
\end{figure}

The sample mean and standard deviation are used to compute a $Z$ statistic in testing $\lambda(a)$, which is reported versus $\lambda$ and $z$, in Figure \ref{fig:poissonz}. A contour plot is used to show the $Z$ value. This statistic seems well behaved, as it appears to be not affected by the configuration in a systematic way. Moreover, $16$ values of $\lambda$ and $100$ values of $z$ were probed, making it likely to satisfy $|\mathrm{Z}| < 3.5$ when the samplers are not troubled.

This last example based on the Poisson distribution should be telling as to why the validation of diagnostic tools in this paper is so limited. Truncation makes statistical higher dimensional and thus more difficult to analyze. Even with a one dimensional base parameter, the truncated distribution would required a rather complicated output to be accurately assessed.

\section{Discussion}\label{sec:discuss}

We have shown that for log-concave distributions it is safer to use Devroye's sampling method rather than the inverse transform. This approach is expected to improve the reliability of statistical analyses that involve both simulation and generalized linear models.

ITS has often been taken for granted to be a safe simulation technique. It is, at least, a very compute-efficient strategy, when $F(x)$ and $q(p)$ tractable. It also offers, with its unmatched generality, an automatic solution that is fast to deploy in practice. DS, instead, looks to be used mostly sparsely, potentially as a patch in cases where no other RS is easy to implement. DS actually represents the automatic counterpart of ITS among log-concave densities, which is why we ship an \texttt{R} package entirely based on it. In RS, a 20-25\% acceptance rate can be improved upon, but not in terms of orders of magnitude, so it is worth considering even if suboptimal, at least from a programming standpoint, since it can address a large set of models in few lines of code.

Our proposal exploits an under-considered aspect of log-concavity under truncation, that is, its closure, which makes DS a scalable sampler even under extreme truncation, wherein other envelopes in RS may work inefficiently. Not all famous log-concave distributions are implemented in the package, but the user may easily add any on their own, though we implemented all famous Tweedie distributions for regression purposes.

Our illustrative examples show that simulation of truncated distributions is currently, at best, a perilous task, since it pushes computation to its limits. We implemented just few distributions, while planning to add more of them in the future, defaults having been tested where present. All the other untested log-concave distributions are left to the user's care who, in the light of above results, uses them at a risk.

There are some open problems, like implementing a sampler for truncated distributions that can work even with un-normalized distributions. This would be much of an improvement, as DS requires $\trunc{f}(x)$ to integrate or summate to unit, so it needs to evaluate the normalization constant involving $F(a)$ and $F(b)$. Other samplers do not require such a knowledge, for instance, \cite{Wild} propose a piecewise-exponential envelope for log-concave distributions that is tailored down to any specific $f(x)$ in an adaptive fashion. This algorithm exploits the fact that a log-concave distribution can not only be upper-bounded, but also lower-bounded, by means of two piecewise-exponential functions, and both help controlling the acceptance rate.

The adaptive sampler can be seen as a more robust alternative to DS in the sense we outlined, as it requires evaluating just $f(x)$ and not $F(x)$, though it is more elaborate. Indeed, DS involves an envelope that is ruled by a location parameter, $m$ and a scale parameter $\sigma=1/f(m)$, which allow for a simple generation of proposed variates and subsequent rejection test. The adaptive sampler relies instead on entirely different envelopes from one target $f(x)$ to another. This latter sampling strategy deserves consideration, but it will be explored further in the future of our \texttt{truncLCdist} package. For now, a rather automatic sampler is provided, whose execution time is bounded in probability, wherein the truncation is tractable, that is, within the safety margin summarized by $\eta$ in the examples.

Possible extensions in the future should definitely target all non-log-concave distributions \cite{Martino_2010}, which we have not covered in our application, due to the lack of a solution of comparable generality. In the case of failure of DS, likely due to numerical instability of $F(x)$, it might be possible to resort to some adaptive version of rejection sampling \cite{Gilks_1992,Wild}. In this case, more acumen in required in coding than it was the focus of this paper. DS needs little tuning, so the comparison with ITS is less viced by the lack of wit in coding. As far as the issues highlighted in this paper look like subtleties, it should be understood why we are careful not to overestimate our own coding skills. Exponential approximations of the normalization constant are also possible in line with Geweke \cite{Geweke_1986} at least for the normal, but also for other log-concave distributions based on heuristic grounds.

DS can be made more safe to a wider extent than currently offered by \texttt{truncLCdist}, just by providing better implementations of $f(x)$, $F(x)$. It is indeed the evaluation of $f(x)$ and of $F(x)$ that troubles the evaluation of $f_\interv(x)$. In the meantime, our package serves as a wrapper that combines together $f(x)$ and $F(x)$ automatically, in order to provide reliable truncated samples.

The \texttt{R} package \texttt{truncLCdist}, used in the simulations for this paper, is publicly available at \verb|https://github.com/mlambardi/truncLCdist|.

\end{document}